\documentclass[%
 reprint,
 amsmath,amssymb,
 aps,
pra,
]{revtex4-1}

\usepackage{graphicx}
\usepackage{dcolumn}
\usepackage{bm}
\usepackage[hidelinks]{hyperref}
\usepackage{textcomp}

\usepackage{color}

\newcommand{\bcl}{{$B_\text{coil}$}}
\newcommand{\epb}{{$\vec{E}\s {\perp}\s\vec{B}$}}

\newcommand{\cmnt}[1]{\ignorespaces}
\newcommand{\s}{{\nobreak\hspace{.2em}}}

\begin{document}

\title{Controlling Spin-flips of Molecules in an Electromagnetic Trap}%

\author{David Reens}
\thanks{dave.reens@colorado.edu.}

\author{Hao Wu}

\author{Tim Langen}%
\altaffiliation{Present Address: 5. Physikalisches Institut and Center for Integrated Quantum Science and Technology (IQST), Universit\"at Stuttgart, Pfaffenwaldring 57, 70569 Stuttgart, Germany}

\author{Jun Ye}
\affiliation{JILA, National Institute of Standards and Technology and the University of Colorado and\\ Department of Physics, University of Colorado, Boulder, Colorado 80309-0440, USA}

\date{\today}

\begin{abstract}
Doubly dipolar molecules exhibit complex internal spin-dynamics when electric and magnetic fields are both applied.
Near magnetic trap minima, these spin-dynamics lead to enhancements in Majorana spin-flip transitions by many orders of magnitude relative to atoms, and are thus an important obstacle for progress in molecule trapping and cooling.
We conclusively demonstrate and address this with OH molecules in a trap geometry where spin-flip losses can be tuned from over $200 \text{ s}^{-1} $ to below our $2\text{ s}^{-1}$ vacuum limited loss rate with only a simple external bias coil and with minimal impact on trap depth and gradient.
\end{abstract}

\maketitle

\section{Introduction}

The ultracold regime extends toward molecules on many fronts~\cite{Carr2009}.
Feshbach molecules at the BEC-BCS crossover have been studied~\cite{Greiner2003, Zwierlein2003, Jochim2003, Bourdel2004}, ground state alkali dimers continue to progress~\cite{Ni2008, Danzl2010, Takekoshi2014, Molony2014, Park2015, Guo2016, Drews2017, Liu2017, Rvachov2017},
and KRb polar molecules have reached quantum degeneracy in an optical lattice~\cite{Moses2015}.
Recently developed laser cooling strategies are tackling certain nearly vibrationally diagonal molecules~\cite{Stuhl2008, Hummon2013, Barry2014, Zhelyazkova2014, Hemmerling2016, Truppe2017}.
A diverse array of alternative strategies have succeeded on other molecules~\cite{Weinstein1998, Bethlem1999, Bochinski2003, Narevicius2008, Wiederkehr2012, Marx2015, Prehn2016, Liu2017a}, including those now enabling molecular collision studies~\cite{Sawyer2011, Zastrow2014, Klein2016, Wu2017}.
Many of these directly cooled molecules will require secondary strategies like evaporation or sympathetic cooling to make further gains in phase space density~\cite{Parazzoli2011, Stuhl2012evap, Quemener2016}.
They also may face a familiar challenge: spin flip loss near the zero of a magnetic trap, but dramatically enhanced for doubly dipolar molecules, relative to atoms, due to their internal spin dynamics in mixed electric and magnetic fields.

Spin flips were directly observed for magnetically trapped atoms near $50\s\mu\text{K}$ and overcome with a time-orbiting potential trap~\cite{Petrich1995} or an optically plugged trap~\cite{Davis1995}, enabling the first production of Bose-Einstein condensates.
Non-laser-based molecular cooling experiments begin at modest temperatures and require trap strengths typically only attained with quadrupole fields~\cite{Weinstein1998, Sawyer2008, Riedel2011, Quintero-Perez2014, Akerman2017}.
In the $2\text{ T/cm}$ magnetic quadrupole used in our previous studies of hydroxyl radicals (OH)~\cite{Stuhl2012evap}, spin-flips should not have had a significant influence until the $\mu$K regime, but the application of electric field changes this.
Electric fields applied to magnetically trapped dipolar species offer interesting opportunities to study anisotropic collisions and quantum chemistry~\cite{Stuhl2013}.
They can also be useful for control over state purity~\cite{Stuhl2012uwave}.
But the electric field can also dramatically enhance spin-flip losses, due to internal spin-dynamics that we corroborate with direct experimental evidence for the first time in the present work.
We achieve this with a novel trap geometry that also allows complete removal of the loss with minimal sacrifice of trap strength.

\begin{figure}[tb]
\includegraphics[width=\linewidth]{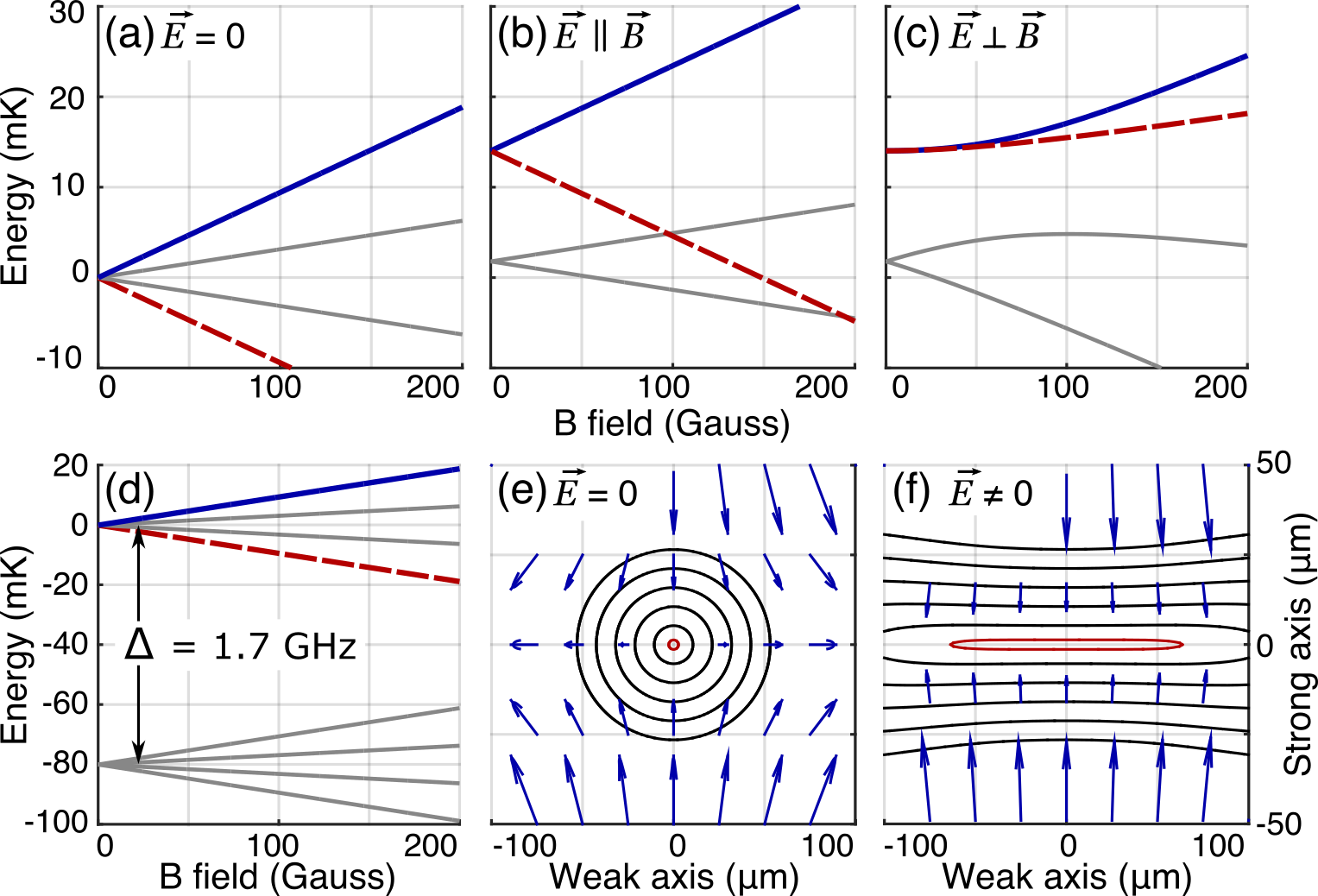}%
\caption{
A uniform electric field, added to magnetically trapped molecules for dipolar studies or other purposes, enhances spin-flip losses.
Note in particular the increased size of the lowest contour (red) in panel (f) relative to panel (e); this result can be understood by considering Zeeman shifts under various conditions as shown in panels (a-c) and described further now.
Four Zeeman split lines in OH's $\mathrm{X}^2\Pi_{3/2}$ manifold are shown\s(a-c), with the doubly stretched state in blue and its spin-flip partner in dashed red.
These states are shown with no electric field\s(a), with $|\vec{E}|\s {=}\s 150\text{ V/cm}$ and $\vec{E}\s $\protect\raisebox{1px}{${\parallel}$}$\s\vec{B}$\s(b), and with \epb{}\s(c).
Note the vastly reduced red-blue splitting in the latter case.
The ground state consists of two parities and four $m$ states each, usually labeled $|\text{parity}\s {=}\s f,e;\;m\s {=}\s{\pm}\s 1/2,{\pm}\s 3/2\rangle$.
The negative parity, electrically strong field seeking manifold sits $\Delta$ below\s(d).
The application of electric field generally drives these parities further apart, but within each parity manifold the exact modification depends on the relative field orientations\s(b-c).
In panels (e-f), energy splitting contours are shown every $40\text{ MHz}$ near the zero of a $2\text{ T/cm}$ magnetic quadrupole trap for OH molecules~\cite{Stuhl2012uwave} with $\vec{E}\s {=}\s 0$\s(e), and with uniform $E\s {=}\s 150\text{ V/cm}$ along the strong axis of the quadrupole\s(f).
The vectors are $d_\text{eff}\vec{E}\s {+}\s\text{sign}(\vec{E}\cdot\vec{B})\mu_\text{eff}\vec{B}$, the proper quantization axis for well-trapped molecules as described in the text.
}
\label{fig:blocking}
\end{figure}

\section{Loss Mechanism\label{sec:mech}}

The internal spin-dynamics leading to spin-flip enhancement are subtle, having eluded two previous investigations:
In Ref.~\cite{Lara2008} the analogues of atomic spin-flip loss for molecules in mixed fields were modeled. 
It was concluded that no significant loss enhancement due to electric field would be evident. However, this conclusion holds only for the approximate Hamiltonian used in that study, not more generally. In Ref.~\cite{Bohn2013} it was correctly noted that Hund's case (a) molecules maintain a quantization axis in mixed fields.
The states of the molecule were shown to align with one of the two quantization axes set by the vector fields $\vec{X}^{\pm}=d_\text{eff}\vec{E} \s {\pm}\s  \mu_\text{eff}\vec{B}$
\s\footnote{The authors use $\mu_\text{eff}\vec{B} \s {\pm}\s  d_\text{eff}\vec{E}$. We reverse this to provide a more physical connection to our experiment, where the electric field is fixed.},
$\mu_\text{eff}$ and $d_\text{eff}$ the effective dipole moments of the molecule in uncombined fields.
The key idea is that Hund's case (a) molecules have both dipole moments fixed to their internuclear axis, so that in the molecular frame, the energy shifts from the two fields combine like vectors. 
It was also shown that the combined Stark-Zeeman energy shift of the molecule is proportional to the length of either $\vec{X}^{\pm}$, depending on the choice of quantization axis, with proportionality given by an $m$ quantum number.
This basis was dubbed ``Hund's Case X," and it was asserted that the existence of the Hund's Case X quantization axis would prevent flips near the zero of a quadrupole trap. 
As we now describe, the loss is actually enhanced, but this Hund's Case X basis actually proves very useful in explaining why.

We begin with an intuitive picture.
In order to remain well trapped in combined fields, a molecule must remain weak field seeking with respect to both fields, i.e. doubly stretched.
This means that the quantization axis to which it aligns should correspond to the vector field $\vec{X}^{\pm}$ with maximal length at the molecule's location.
In a geometry where the fields are continuously rotating, the maximal length vector field can be either of $d_\text{eff}\vec{E} \s {\pm}\s  \mu_\text{eff}\vec{B}$, depending on whether the fields are oriented closer to parallel or antiparallel.
Consider a molecule in a magnetic quadrupole trap with a uniform electric field.
This trap has two hemispheres, a parallel hemisphere where the fields are closer to parallel, and vice versa.
The hemispheres are separated by a plane where \epb{}, which intersects the trap center.
Now suppose a molecule in the doubly stretched state begins in the parallel hemisphere.
To be doubly stretched, it must be aligned to the sum quantization axis set by the vector field $\vec{X}^{+} = d_\text{eff}\vec{E} \s {+}\s  \mu_\text{eff}\vec{B}$.
If its trajectory carries it near the trap center where the magnetic field is small, the electric field does indeed maintain the quantization axis and the molecule's alignment with it.
However when the molecule enters the antiparallel hemisphere, the magnitude of this quantization axis now decreases with increasing magnetic field. This molecule has therefore spin-flipped from the doubly stretched state to a magnetically strong field seeking state. These states are degenerate in the plane where \epb{}, leading to loss.

This intuition agrees with a more rigorous analysis of the energy splitting $G$ between the trapped state and its spin-flip partner.
By diagonalizing the approximate eight state ground molecular Hamiltonian for OH, subtracting the relevant state energies and Taylor expanding, we find:
\begin{equation}
\label{eqn:energetics}
G(\mathcal{B}_\perp,\mathcal{B}_{||},\mathcal{E}) = 2\mathcal{B}_{||} + 4.3\!\cdot\!\mathcal{B}_\perp^3\frac{\Delta^2}{\mathcal{E}^4} + \mathcal{O}(\mathcal{B}_{||}^2,\mathcal{B}_\perp^4)
\end{equation}
Here $\mathcal{B}_{\perp,||}\s {=}\s\mu_\text{eff}\vec{B}\cdot\hat{e}_{\perp,||}$, where $\hat{e}$ is the unit vector in the labeled direction relative to the electric field.
$\mathcal{E}\s {=}\s |d_\text{eff}\vec{E}|$, $\Delta$ is the lambda doubling (see Fig.~\ref{fig:blocking}d).
The relevant splitting is not quite zero where \epb{} and $B_{||}\s {=}\s 0$ thanks to $\Delta$, but nonetheless reaches a deep minimum; the remaining Zeeman splitting is reduced from linear to cubic in magnetic field (Fig.~\ref{fig:blocking}).
This Zeeman splitting suppression is in fact a known phenomenon in the precision measurement community~\cite{Player1970,Hudson2002},
and experimentalists have exploited it to suppress the influence of magnetic fields in electron EDM measurements.
However, in the case of applying mixed fields during trapping, this suppression is not beneficial but rather detrimental.

\newcommand{\shiftright}[2]{\makebox[#1][r]{\makebox[0pt][l]{#2}}}
\begin{table}[t]
\caption{
Enhancements ($\eta$) and loss rates ($\gamma$) for OH with typical applied fields.
Zero field values are equivalent to traditional spin-flip loss.
Electric field is required during evaporation and spectroscopy to open avoided crossings~\cite{Stuhl2012evap,Stuhl2012uwave}, or applied to polarize the molecules and study collisions~\cite{Stuhl2013}.
}
\label{tab:rates}
\begin{tabular*}{\linewidth}{l*{4}{@{\quad}c}@{\extracolsep{\fill}}l}
\hline\hline
 & \raisebox{-1.3ex}{\shiftright{4pt}{55 mK}} & & \raisebox{-1.3ex}{\shiftright{4pt}{5 mK}} & & \\
\raisebox{1.5ex}{$E$ (V/cm)} & $\eta$ & $\gamma\s (s^{-1})$ & $\eta$ & $\gamma\s (s^{-1})$ & \raisebox{1.5ex}{Purpose} \\
\hline
0 		&1 		& 0.02 	& 1 		& 1.3 	& Zero Field \\
300 		& 5 		& 0.1 	& 9 		& 11 		& Evaporation \\
550 		& 17 		& 0.3 	& 40 		& 50 		& Spectroscopy \\
3000 	& 1000 	& 19 		& 1600 	& 2000 	& Polarizing \\
\hline\hline
\end{tabular*}
\end{table}

To deduce the effect of this loss plane on the ensemble, we consider molecular trajectories in light of the Landau Zener formula:
\begin{equation}
\label{eqn:lz}
P_\text{hop}=e^{-\pi\kappa^2/2\hbar\dot{G}},
\end{equation}
which relates the probability of diabatically hopping between two states $P_\text{hop}$ to their energetic coupling $\kappa$ and their rate of approach $\dot{G}\s {=}\s v_zdG/dz$.
Here $z$ and $v_z$ are normal to the \epb{} plane, and we neglect the components of $\dot{G}$ due to the other coordinates since from Eqn.~\ref{eqn:energetics} it is clear that $G$ grows predominately in one direction.
We can also set $\kappa$ to the minimum energy gap along the trajectory, which is found in the plane.
This facilitates direct numerical computation of the loss rate ($\gamma$) by integrating the molecule flux through the plane for a thermal distribution, weighted by the hopping probability. See Eqn.~\ref{eq:fullInt} in App.~\ref{sec:eic} for the full expression.
We perform these integrations for OH over the velocity distribution in a $2\text{ T/cm}$ magnetic quadrupole~\cite{Sawyer2008} under various electric fields in Tab.~\ref{tab:rates}.

We have also developed an algebraic scaling law, which yields the electric field induced loss enhancement factor
\begin{equation}
\label{eq:etaMT}
\eta=\frac{3}{11} \left(\frac{d_\text{eff}E}{\sqrt{\kappa\Delta}}\right)^{8/3},
\end{equation}
see App.~\ref{sec:der} for the full derivation.
Here $\kappa$ represents a characteristic energy scale for spin-flips that can be derived by setting $P_\text{hop}\s {=}1/e$ in Eqn.~\ref{eqn:lz} and using a typical value of $v_z$.
This means that for electric fields with $d_\text{eff}E>\sqrt{\kappa\Delta}$, the loss enhancement is almost cubic with electric field.
Crucially, it is not $\Delta$ that sets the relevant scale, as one might naively suppose given that this is the energy beyond which the Stark effect is linear and the molecule is polarized.
Instead it is $\sqrt{\kappa\Delta}$, which is in general much smaller; $\kappa\s {=}\s 5\text{ MHz}$ for OH in our trap, while $\Delta\s {=}\s1.7\text{ GHz}$.

Returning to the numerical approach, the direct integration of flux is a key improvement relative to our previous work~\cite{Stuhl2013}, where electric fields were applied to study collisions.
The mechanism of molecular spin-flip loss was identified, and an attempt was made to deconvolve it from the collisional effect of the electric field.
Revisiting this with the direct integration of flux, we find a three-fold larger loss magnitude, enough to explain a significant portion of the effect previously attributed to collisions, see App.~\ref{sec:eic}.
In light of this, it becomes especially important to perform direct, unconvolved experimental verification of both the magnitude of the loss effect and the validity of our loss-flux calculations.
We now present the new trap where this is achieved.

\section{Experiments and Results\label{sec:results}}

\begin{figure}[tb]
\includegraphics[width=\linewidth]{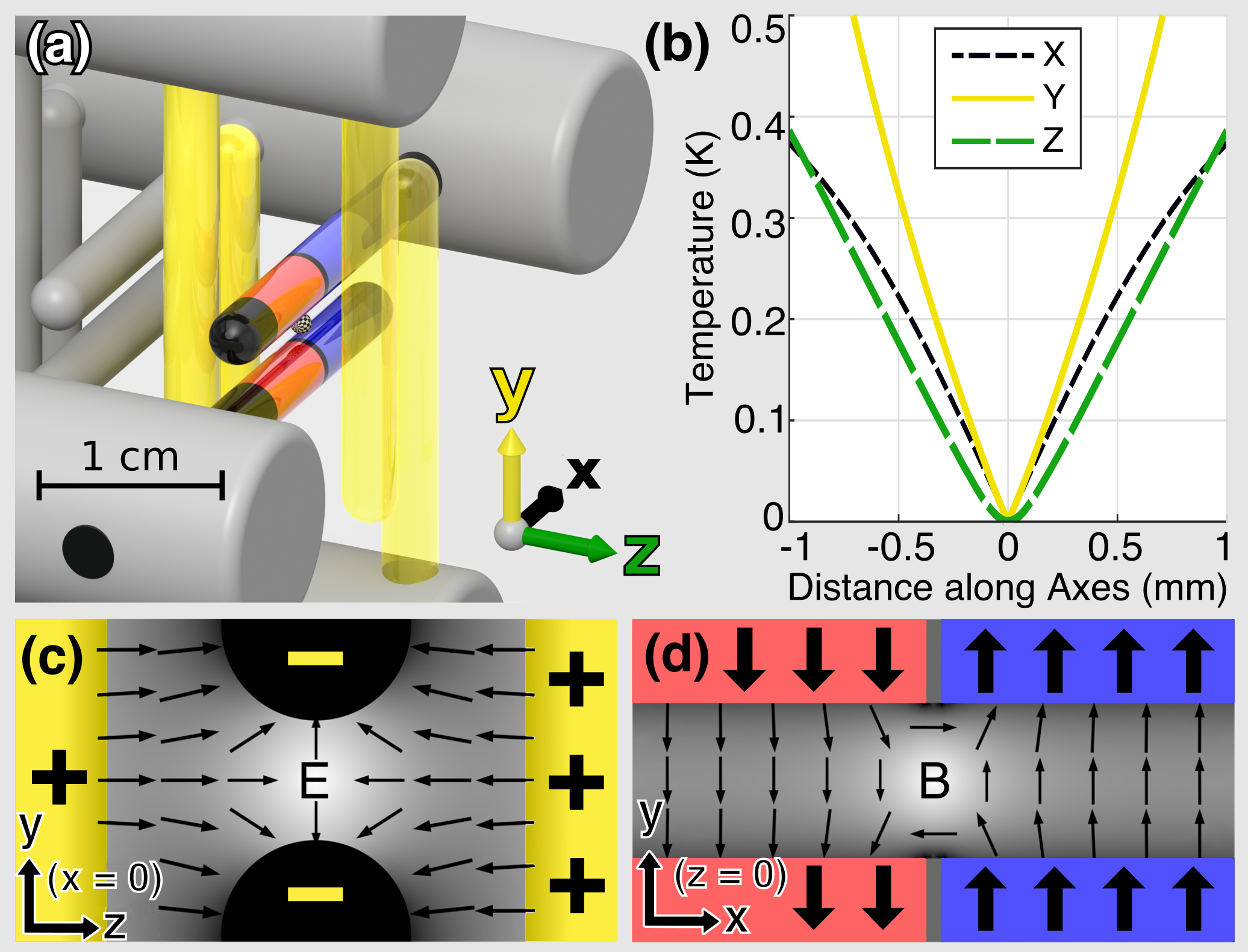}
\caption{
The last six pins of our Stark decelerator~\cite{Sawyer2008} form the trap\s(a), which is $0.45\text{ K}$ deep with trap frequency $\nu\s {\approx}\s 4\text{ kHz}$\s(b).
Along $y$ the trap is bounded by the $2\text{ mm}$ pin spacing.
The yellow pins are positively charged and the central pin pair negatively, which forms a 2D electric quadrupole trap with zero along the $x$-axis.
This is shown for the $x\s {=}\s 0$ plane\s(c), with yellow pins artificially projected for clarity since they don't actually intersect the plane.
The central pins are magnetized, with two domains each.
Blue indicates magnetization along $+\hat{y}$, red along $-\hat{y}$.
These domains produce a magnetic quadrupole trap with zero along the $z$-axis, shown in the $z\s {=}\s 0$ plane\s(d).
\label{fig:CAD}}
\end{figure}

Our idea is to use a pair of 2D quadrupole traps, one magnetic and the other electric, with orthogonal centerlines (Fig.~\ref{fig:CAD}):
\begin{equation}
\label{eqn:BE}
\vec{B}=B'x\hat{y}-B'y\hat{x}\quad\quad\vec{E}=E'y\hat{y}-E'z\hat{z}
\end{equation}
We achieve these fields in a geometry that matches our Stark decelerator~\cite{Bochinski2003}.
This geometry features large spin-flip loss, since \epb{} in both the $x\s {=}\s 0$ and $y\s {=}\s 0$ planes, and from Eqn.~\ref{eqn:energetics}, $G\s {=} \s \mathcal{B}_\perp^3\Delta^2/\mathcal{E}^4$ will be generally very small due to the large $\mathcal{E}$.
However, by adding a small magnetic field $\vec{B}\s {=}\s B_\text{coil}\hat{z}$ along the centerline of the magnetic quadrupole with an external bias coil, a dramatic change can be made to the surfaces where \epb{} with minimal perturbation of the trapping potential.

\begin{figure}[tb]
\includegraphics[width=\linewidth]{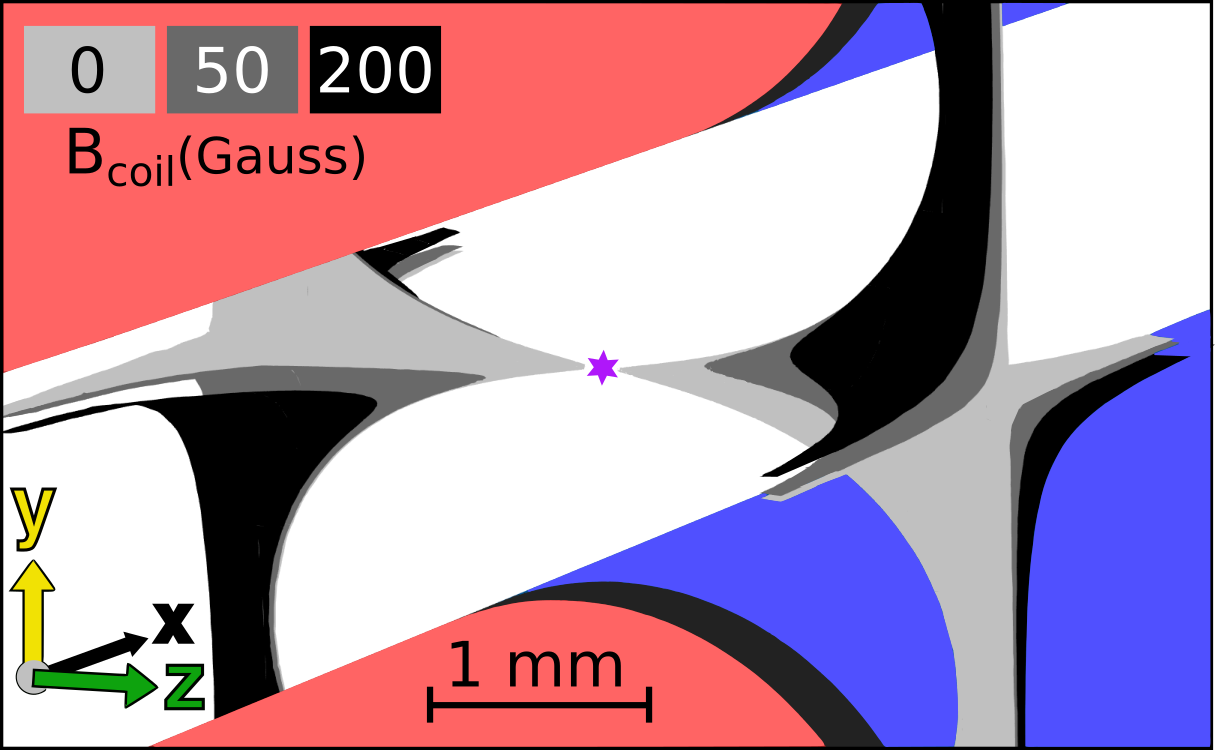}%
\caption{
Surfaces where spin-flips can occur are shown for three values of \bcl{} in light gray, dark gray, and black.
The magnetic pins are shown as in Fig.~\ref{fig:CAD} for context.
The purple star marks the trap center, to which molecules are confined within a \protect\raisebox{2.5px}{\texttildelow} $\!\!1\text{ mm}$ diameter.
\label{fig:LSurfs}}
\end{figure}

\begin{figure}[tb]
\includegraphics[width=\linewidth]{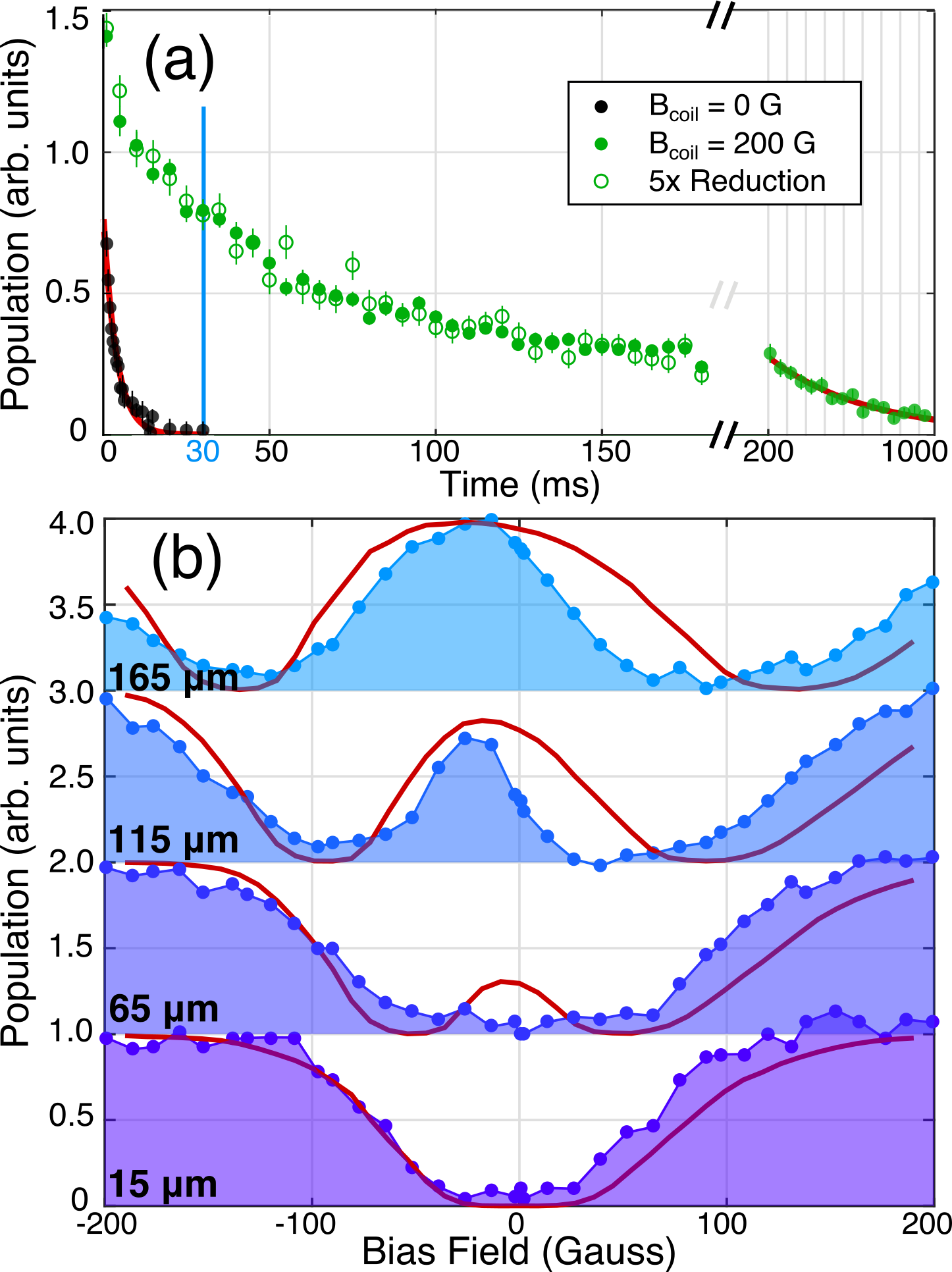}%
\caption{
Time traces\s(a) without bias field\s(black), with bias field\s(green dots), and with modulated density\s(green circles).
One body fits\s(red) give loss rates of $200\text{ s}^{-1}$ without bias field and $2\text{ s}^{-1}$ with full bias field at long times, in agreement with our background gas pressure.
At the fixed time $30\text{ ms}$, population is shown as a function of both pin translation and bias field\s(b), for several values of pin translation, labeled relative to perfect alignment.
Fits\s(red) are calculated by integrating the molecule flux of a thermal ensemble through surfaces where \epb.
\label{fig:WVplot}}
\end{figure}

\bcl{} morphs the \epb{} surface from a pair of planes into the hyperbolic sheet given by
$x\s {\cdot}\s y\s {=}\s  z\s {\cdot}\s B_\text{coil}/B'$
(Substitute Eqn.~\ref{eqn:BE} into  $\vec{E}\s {\cdot}\s\vec{B}\s {=}\s 0$).
This means that \epb{} is pushed away from the $z$-axis where $\vec{B}$ is smallest.
In Fig.~\ref{fig:LSurfs}, the surfaces where \epb{} for several \bcl{} magnitudes are calculated and shown wherever $G\s {\le}\s\kappa$.
The loss regions ought to be tuned far enough from the trap center that molecules cannot access them.
This is indeed what we observe, note the striking difference in trap lifetimes in Fig.~\ref{fig:WVplot}a.
With only $200\s\text{G}$ bias field (the trap is $5\s\text{kG}$ deep) the loss is suppressed below that due to background gas.

To further verify our understandings of the loss mechanism, we translated one of the magnetic pins along $\hat{x}$.
This pin translation disrupts the idealized 2D magnetic quadrupole by adding a small trapping field $\vec{B} \s {\propto}\s  B^\prime z\hat{z}$, which significantly alters the topology of the \epb{} surface and the overall loss rate in the trap. We also compute loss rates for all values of pin translation and \bcl{} by numerically integrating the loss flux through these unusual loss surfaces via the Landau-Zener formula, just as for the simpler quadrupole geometry discussed previously.  The calculated populations after $30\text{ ms}$ in the trap have a reasonable agreement with the measurements (Fig.~\ref{fig:WVplot}b).
The direct integral calculation uses only the temperature of a purely thermal distribution as a free parameter, and does not involve computation of any trajectories.
The temperature fits to $170\s {\pm}\s 20\s\text{mK}$\s\footnote{Calculation performed in COMSOL: \href{https://github.com/dreens/spin-flip-integration/}{Source Code}}.
An intuitive explanation for the intriguing double well structure in Fig.~\ref{fig:blocking} is that \bcl{} first translates the magnetic zero along the z-axis, overlapping it with larger electric fields at first before moving it out of the trap.

With strong experimental confirmation of the molecular spin-flip loss enhancement, we can move on to generalize beyond OH.
Hund's case (a) states are most susceptible in the sense that smaller electric fields are sufficient to cause a significant problem, but with enough electric field any state exhibiting competition between electric and magnetic fields for alignment of the molecule or atom will be susceptible.
One way to avoid competition is for the fields to couple to unrelated parts of the Hamiltonian, which happens to a limited extent for Hund's case (b) states without electron orbital angular momentum ($\Sigma$ states, $\Lambda\s {=}\s 0$)~\cite{Bohn2013}.
In these states, which include most laser-cooled molecules thus far, the electric and magnetic fields couple to rotation and spin respectively, which are only related by the spin-rotation coupling constant.
This constant is usually in the tens of MHz~\cite{Quemener2016}, so molecular spin-flip loss remains quite significant.
The inclusion of hyperfine requires a careful case-by-case investigation. For OH, it would initially seem to add an extra splitting that could protect from spin-flips, but in fact the loss plane is only shifted slightly away from \epb{} and retains the same area.
For YO\s\footnote{This is particularly relevant given the recently realized 3D MOT for YO}, certain hyperfine states can avoid spin-flip loss entirely when electric fields are applied.
These states are characterized by significant electron-spin-to-nuclear-spin dipolar coupling, which results in a protective gap regardless of field orientation.

It is also instructive to consider the related case of a pure electrostatic trap.
Here there is always some zero field parity splitting that prevents the orientation-reversing spin flips we have been discussing.
However, this same splitting pushes all states with the same sign of $m$, the field alignment quantum number, very close to one another, leading to loss via Landau-Zener transitions other than the $m$ to $-m$ spin-flip~\cite{Wall2010}.
Intriguingly, the addition of a homogeneous magnetic field can actually suppress this loss~\cite{Meek2011}.

The present trap, in addition to providing the desired experimental testing ground for molecular spin-flip loss, produces large $5\s\text{T/cm}$ trap gradients useful for maintaining high densities to facilitate collisional studies.
This is in contrast with other strategies for plugging the hole of a magnetic trap which often lead to a reduction in trap gradient.
With loss removed, we observe a population trend whose initially fast decay rate decreases over time (Fig.~\ref{fig:WVplot}a, green dots), suggesting a two-body collisional effect.
We test this by reducing the initial population fivefold but without changing its spatial or velocity distribution, and then scale the resulting trend by five (green circles).
This technique is described further toward the end of App.~\ref{sec:evap}.
If collisions had contributed, this new trend would show less decay, but we observe no significant change.
This seeming lack of collisions is likely due to the much higher initial temperature of $170\s\text{mK}$, in contrast to the earlier work at and below $50\s\text{mK}$~\cite{Stuhl2012evap}. An alternative hypothesis for the population trend is the existence of chaotic trap orbits with long escape times~\cite{Gonzalez-Ferez2014}.  The understanding of electric field enhanced spin-flip loss brings an important consequence to the use of RF knife under electric field employed in the forced evaporation, especially at low temperatures. We present the effect of evaporation at intermediate temperatures (${\sim}\,30\text{ mK}$) in App.~\ref{sec:evap}.  Moving forward, we aim to increase the density by means of several improvements~\cite{Even2015,Segev2017}.

\section{Conclusions}

Molecule enhanced spin-flip loss arises in mixed electric and magnetic fields due to a competition between field quantization axes.
We conclusively demonstrate and suppress this effect using our dual magnetic and electric quadrupole trap, which is also an ideal setting for further progress in collisional physics thanks to its large trap gradient.
Our calculation of the magnitude of spin-flip loss via flux through surfaces where \epb{} enables detailed predictions of how its location and magnitude ought to scale with bias field and trap alignment, which we experimentally verify.
Our results correct existing predictions about molecular spin-flips in mixed fields and pave the way toward further improvements in molecule trapping and cooling.

\begin{acknowledgments}
We acknowledge the Gordon and Betty Moore Foundation, the ARO-MURI, JILA PFC, and NIST for their financial support.
T.L. acknowledges support from the Alexander von Humboldt Foundation through a Feodor Lynen Fellowship.
We thank J.L. Bohn, S.Y.T. van de Meerakker, and M.T. Hummon for helpful discussions.
We thank Goulven Qu\'em\'ener for his continued involvement in this research.
\end{acknowledgments}

\section*{Appendices}

The present study on the role of mixed fields for spin-flip loss evolved out of our continuing investigations into the collisional processes of trapped OH molecules in a magnetic quadrupole trap reported in Refs.~\cite{Stuhl2013,Stuhl2012evap}. The current investigations have revealed that the spin-flip loss can be substantially enhanced when an electric field is applied to the magnetic trap, and thus an important fraction of the inelastic collisional loss under various electric fields is in fact attributable to spin-flip losses. In App.~\ref{sec:eic} and App.~\ref{sec:evap} we provide further information on the electric field-induced trap loss and evaporative cooling, respectively. In App.~\ref{sec:der} we provide an algebraic derivation of the loss enhancement factor presented in Eqn.~\ref{eq:etaMT}.

\appendix

\section{Electric Field-Induced Trap Losses\label{sec:eic}}

Ref.~\cite{Stuhl2013} introduced the single particle spin-flip loss enhancement process and deconvoluted its effect from the inelastic collisional effect (Appendix~A, Ref.~\cite{Stuhl2013}). Since that time, new and more systematic experimental observations have prompted improvements to the analysis that was presented there.

Relative to the previous approach, we make the same simplifying assumptions: loss only occurs in the \epb{} plane, and only the velocity orthogonal to this plane matters as molecules cross this loss plane, and the in-trap population follows a thermalized Maxwell-Boltzmann distribution.
Our improvement relates to the next step, where an integral calculation for the loss rate is performed.
In Ref.~\cite{Stuhl2013} the integration spans the entire 3D spatial distribution, weighted by the frequency of crossing of the center plane and the chance of loss for each crossing:
\begin{equation}
\gamma=\int\displaylimits_0^\infty\!4\pi r^2\,n(r)dr\!\int\displaylimits_0^\infty \!n(v_\theta)dv_\theta\left( \frac{v_\theta}{\pi r} P_\text{hop}(r,v_\theta)\right).
\end{equation}
Here $n(r)$ is the radial distribution function, constrained to satisfy $\int_0^\infty 4\pi r^2n(r)=1$, and of the form $n(r)\propto e^{-\mu_BB' r/kT}$. Likewise $n(v_\theta)$ is the usual normalized Maxwellian velocity distribution.  Implicit in this integration is the assumption that molecules at a given radius $r$ cross the center plane with a frequency of $v_\theta/\pi r$.
This approximation is rather simplistic given that molecules are typically not following circular orbits of constant $v_\theta$ but are in general following some complex trap motion. In addition, the trap is approximated as spherically symmetric to avoid the complication of elliptical coordinates in the three dimension.

A more accurate treatment that we use here is to perform an integration of flux through the loss plane directly:
\begin{equation}
\label{eq:fullInt}
\gamma=\int\displaylimits_0^\infty\!2\pi r \,n(r)dr\!\int\displaylimits_0^\infty\! n(v_z)dv_z\left(v_z P_\text{hop}(r,v_z)\right).
\end{equation}
Here the spatial integration is over the central plane only, hence the $2\pi r$ Jacobean, and the hopping probability is multiplied by $v_z$ to give a flux. The population distribution $n(r)$ is now normalized correctly for an oblate ellipsoidal quadrupole trap, which no longer requires elliptical coordinates since the integration is only in one plane. We change to cylindrical coordinates to highlight our focus on the central plane.
This flux integral gives the desired loss rate without any approximations about molecule orbits or plane-crossing frequency. This rigorous treatment provides precisely an overall scaling factor of $\pi$ relative to the previous estimate. Comparing the integrands and Jacobeans of Eqn. 1 and 2 gives a factor of $\pi$/2, and the change of integration from the spherical trap volume to the loss plane provides a factor of 2 via the distribution $n(r)$.

\begin{figure}[t]
\includegraphics[width=\linewidth]{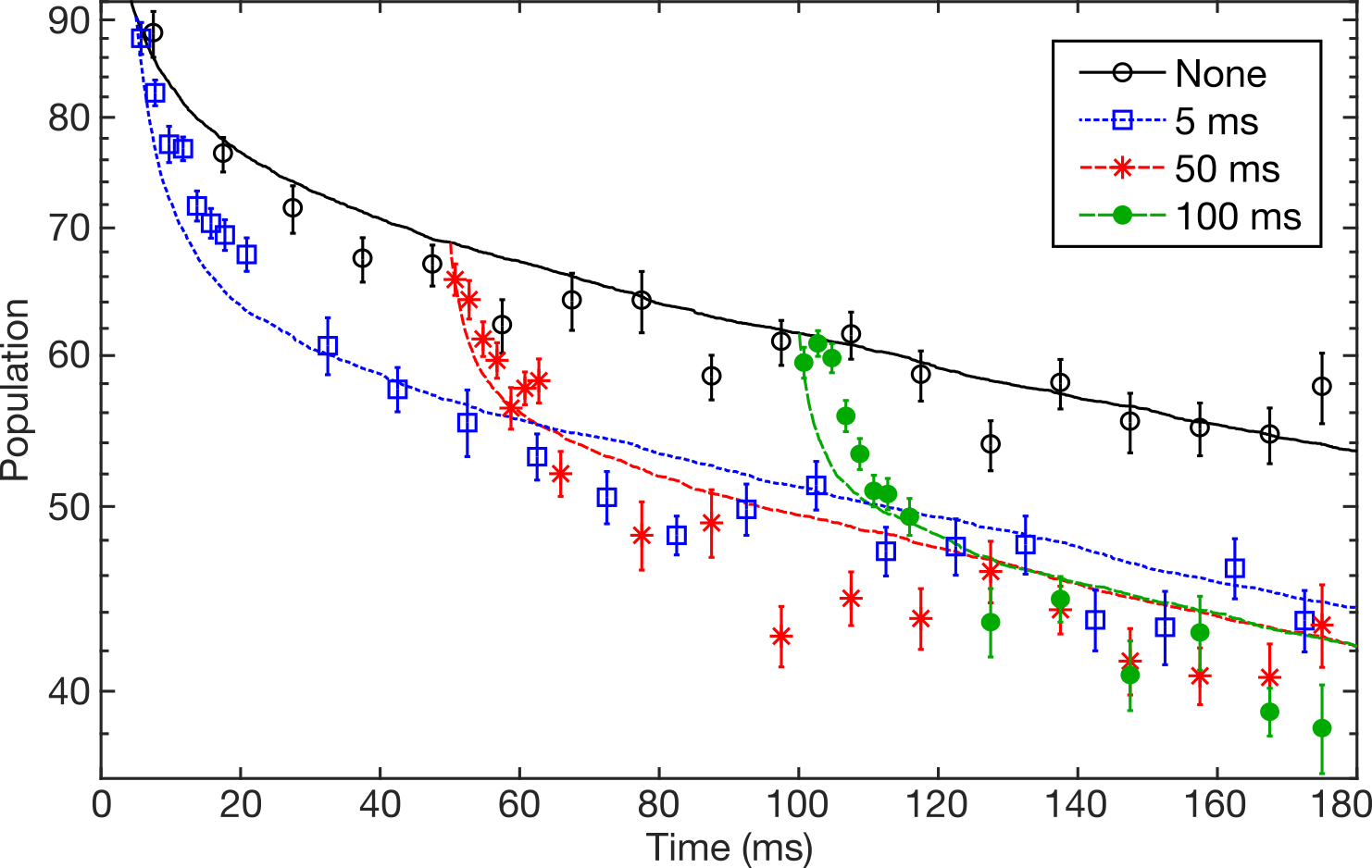}
\caption{Experimental data on electric field-induced loss with an attempted overlap to spin-flip loss simulations. The case of no electric field (black, solid, circles) is compared to electric fields of $3\text{ kV/cm}$ turned on after a wait time indicated in the legend.
\label{fig:eil}}
\end{figure}

\begin{figure}[t]
\includegraphics[width=\linewidth]{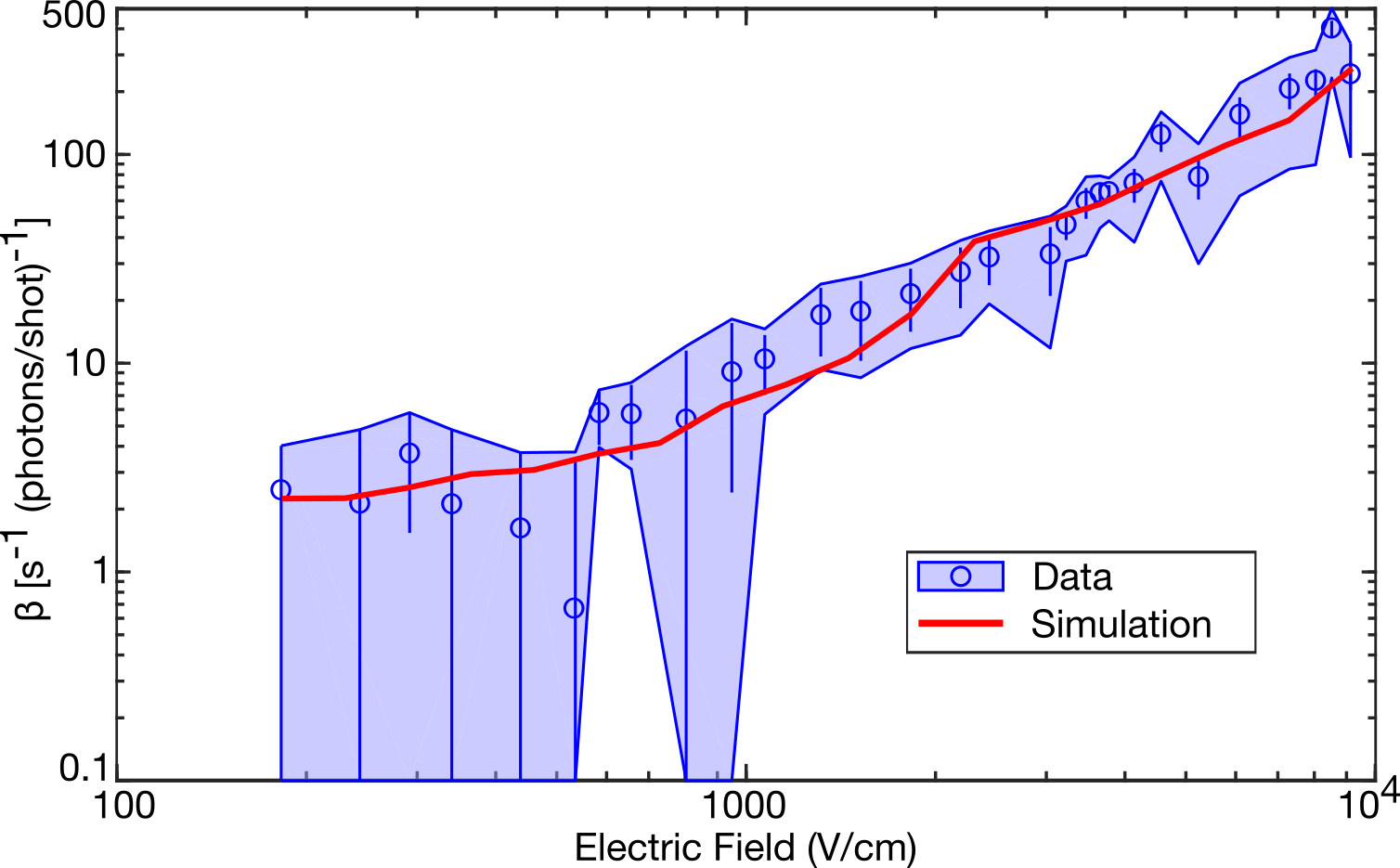}
\caption{Two body fits from~\cite{Stuhl2013} to experimental data like that in Fig.~\ref{fig:eil} but at various electric fields. The blue data points and shaded region are repeated from Fig.~3 of~\cite{Stuhl2013}, where the shading indicates the variation that would be brought about by two-fold changes in $\Gamma$ from spin-flip losses. With a factor of 3 correction noted in this study, the spin flip loss simulation (thick red line) matches the original data within errorbars.\label{fig:beta}}
\end{figure}

The influence of this factor on the deconvolution procedure relates to the details of the two-body fitting routine.
One plus two body fits $\dot{N}=-\Gamma N-\beta N^2$ were performed to various decay trap curves, with the one body rate $\Gamma$ fixed to the value expected due to vacuum scattering and spin-flip loss.
An example of such decay curves is shown in Fig.~\ref{fig:eil}, where electric field is turned on suddenly after various hold times, which is motivated by the desire to vary the trapped sample density.
With the stronger spin-flip loss, we also consider its effect beyond a pure one-body decay. Molecules whose orbits regularly intersect the loss region are lost, after which thermalization would be required to repopulate the loss prone trajectories of phase space. If thermalization is slow, spin-flip loss can have a rate that decreases over time, producing a time dependence of population like that of a two-body effect.
Even though the possibility of a factor of two error in the calculated magnitude of spin-flip loss was considered in Ref.~\cite{Stuhl2013} (Repeated here in Fig.~\ref{fig:beta}, shaded regions), the possibility of its influencing the data in a non-single-particle manner was not addressed. We note however that both the previous and current derivations of spin-flip loss assume a thermal distribution in the trap.

We have performed single particle Monte Carlo simulations of spin-flip loss to further investigate this effect, and we obtain curves such as shown with the experimental data in Fig.~\ref{fig:eil}.
We also performed the same one- plus two-body fitting procedure to the single particle spin-flip loss simulation traces, which yield two-body values that overlap with those derived in Ref.~\cite{Stuhl2013}, see Fig.~\ref{fig:beta}.  This suggests that the spin-flip loss plays an important role in the observed loss data under applied electric fields, and the effect of inelastic collisions is marginal within the errorbars. Still, as we did not involve inelastic collisions in the simulation, there are notable discrepancies between simulation and data, such as in the initial rate of the decays in Fig.~\ref{fig:eil}.
One avenue to try and improve agreement would be to incorporate collisions in the simulation. There are many challenges in the quantitative application of these simulations, such as the existence of various partially trapped substates. The best path forward is to perform future collisional experiments with the single-particle effect removed.

\section{Evaporation\label{sec:evap}}

Ref.~\cite{Stuhl2012evap} describes the processes of evaporation from a magnetic quadrupole trap and of depletion spectroscopy to measure the trap distribution. Both processes require two steps.
First, molecules are transferred from the positive to the negative parity state by applying short pulses of microwaves tuned to a specific range of magnetic fields.
After this transfer, the molecules are still trapped, and only by the subsequent application of a DC electric field to open avoided crossings can these opposite parity molecules escape from the trap.
In the case of using depletion spectroscopy for thermometry, the microwave pulses transfer only a small fraction of population between the opposite parity state to reflect the original population distribution~\cite{Stuhl2012uwave}. A final step is necessary to measure the overall population in the trap by laser induced fluorescence.
The crossings opened by electric field would only allow molecules in the upper $90\%$ of the trap to escape, so it was assumed that a cold population insensitive to the spectroscopic technique would be building up in both parity states at low magnetic fields. Given the existence of spin-flip loss caused by the electric field at the center of the trap where such a cold population would build up, and given its strength for the relevant temperatures and electric fields (Tab.~\ref{tab:rates}), the assumption of cold samples building up in the lower parity state must be reexamined.

\begin{figure}[tb]
\includegraphics[width=8.2cm]{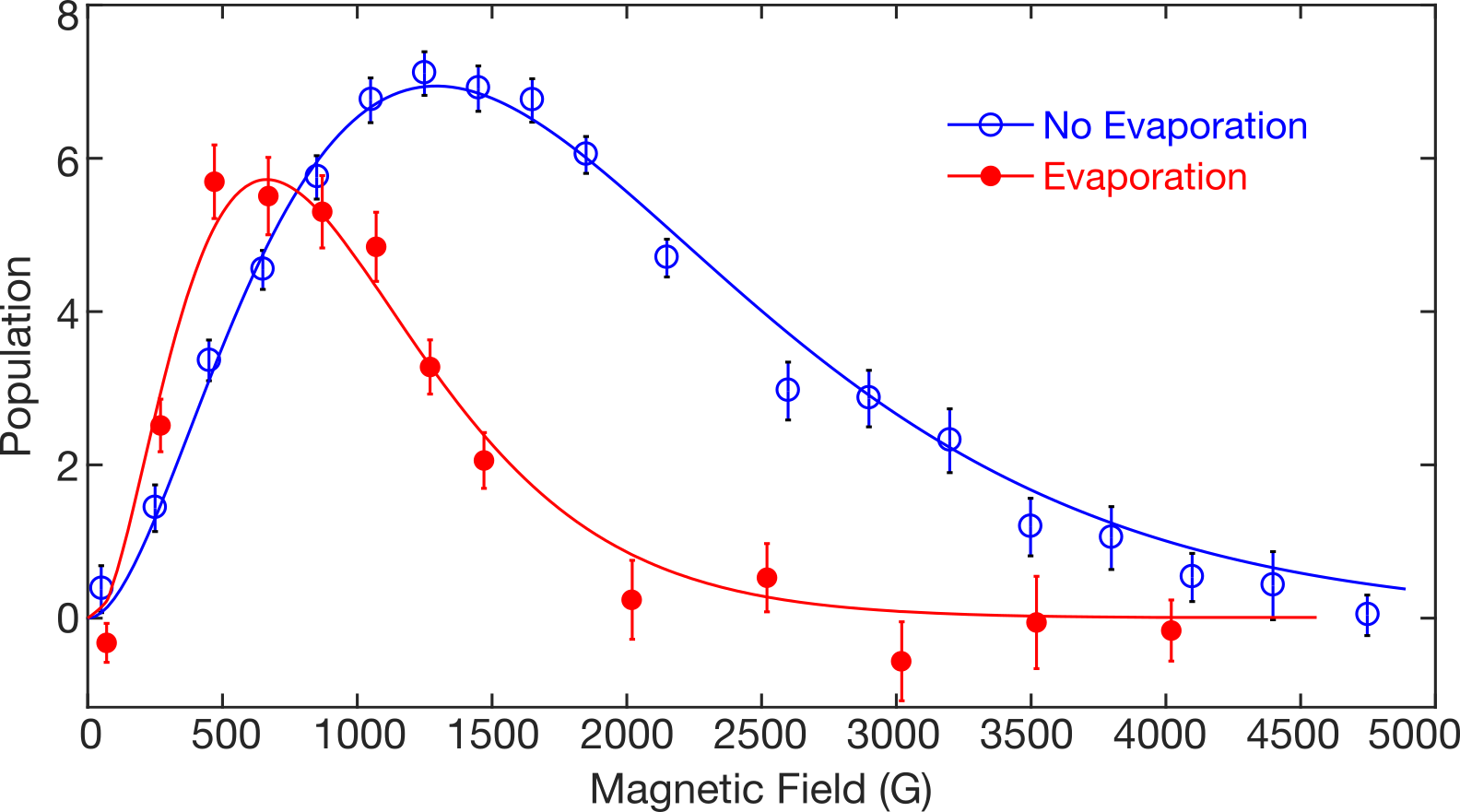}%
\caption{
Depletion spectroscopy spectra are obtained after evaporation (red filled circles) and without evaporation (blue open circles). The integrated areas under the curves correspond to the total number of molecules detected. Solid lines are fits to Maxwell-Boltzmann distributions with temperatures of $59\pm2$~mK without evaporation (blue), and $30\pm2$~mK for evaporation (red). The evaporation achieves a clear density enhancement in the vicinity of $500\text{ G}$.
}
\label{fig:normenhance}
\end{figure}

Some of the temperature fits performed in Fig.~3 of Ref.~\cite{Stuhl2012evap} relied on this assumption, which we now no longer use. We rely on only the directly experimentally accessible spectra, such as those shown in panels (a-c) of Fig.~3 of Ref.~\cite{Stuhl2012evap}. After taking similar measurements repeatedly, the depletion spectra are found to be useful to identify enhancements in density caused by the evaporation. Figure~\ref{fig:normenhance} show such enhancements for evaporation sequences designed to achieve a twofold temperature reduction. The initial temperature of $59\pm2\text{ mK}$ is higher than reported in Ref.~\cite{Stuhl2012evap}, mostly due to a subtle correction to the molecular Hamiltonian.  A detailed calculation of this correction including nearly one hundred ground and excited hyperfine levels is given in Ref.~\cite{Maeda2015}.

Since depletion spectroscopy transfers only a fraction of molecules to the lower parity state at a specific magnetic field value, we integrate the total area enclosed by the spectroscopy curve, which is scaled according to the observed total population by laser induced fluorescence.  
This is necessary because depletion spectroscopy is performed with a train of short microwave pulses lasting over a total time of about a quarter of a trap oscillation, so that molecules are not at all frozen in place. 
Relative to a very brief spectroscopy pulse that would only deplete molecules in a given region at that particular instant, the use of a train of pulses over a longer period of time allows us to sample molecules more widely to boost the signal to noise ratio of spectroscopy. 
The spectroscopy gives a value that is proportional to the true instantaneous population in a specific magnetic field region, but with a scaling factor that allows the signal to be constrained with the measured total number of molecules in the trap. 
We carefully include all of the steps in the error analysis leading to the error bars shown in Fig.~\ref{fig:normenhance}.

In addition to the spatial density enhancement in the low magnetic field region, we can also examine the phase space density (PSD) in the trap, under the assumption of good thermal equilibrium. 
As the population $N$ was reduced by $60\%$ after evaporation, the temperature came down by a factor of $2$. 
For a quadrupole trap $\text{PSD}\,{\propto}\,N\,T^{-9/2}$, which gives a PSD increase of a factor of $10$. 
It is possible for truncation effects to explain a fraction of this effect.
When we perform Maxwell-Boltzmann fits to truncated distribution models, we can see PSD enhancements of at most $6$, at which point the truncated models no longer bear much resemblance to a thermal distribution and cannot be fit.

We have also performed another independent verification of the collisional effect by comparing the populations under two closely related experimental sequences.
The first is a normal evaporation sequence and the second is identical but with a time-reversed microwave frequency chirp, so that the population cut goes backwards from deep to shallow in the trap.
This comparison subjects all molecules to the same integrated microwave power, and thus the two conditions should be equivalent in a situation with only single particle effects.
With respect to collisional effects, the time-reversed case functions like a truncation, preventing molecules that would otherwise have collisionally thermalized to lower temperatures from doing so.
To whatever extent an evaporation is successful in facilitating beneficial thermalizing collisions, the time-reversed condition should yield fewer molecules.
We consistently observe this at the $(6\,{\pm}\,2)\%$ level, pointing to an evaporative effect despite the negative influence of spin-flip losses.
We have also experimentally observed that under less ideal initial conditions, such as higher initial temperatures of $80\text{ mK}$ resulting from poorer decelerator performance, the density enhancements at low magnetic field and the forward backward differences both disappear, confirming the role of evaporation. 

Moving forward, we can now use the newly developed capability of reducing the population without perturbing its phase space distribution, as mentioned above in Sec.~\ref{sec:results}.
This ought to reduce the influence of collisional processes, but keep any single particle effects the same, thus disambiguating the two.
Many possible approaches have key drawbacks, for example changing the partial pressure of water in our supersonic expansion would require changing the temperature of the valve and thereby influencing the initial speed of the beam. 
We opt for the application of microwaves during deceleration, leading to a probability for transitioning from a weak to strong field seeking state and being deflected out of the beam.
We tune the microwaves to be resonant only at low magnitudes of electric fields, experienced by all molecules when flying through a de-energized stage just after switching.
The microwaves are applied via horn and have a $17\text{ cm}$ wavelength, so that microwave power variations across the cloud are minimal.
The microwaves are applied early during deceleration, so that the molecules have many stages of deceleration left to remix any outstanding asymmetries in the removal process.
It is difficult to experimentally verify that the phase space distribution is truly unaffected, but in one projection of phase space, the time of flight profile of slowed molecules after deceleration, the distribution seems to be unaffected even by tenfold reductions using this technique.

While the role of collisional effects in Ref.~\cite{Stuhl2012evap} is reduced by spin-flip losses, especially at low temperatures below 10 mK, spectroscopic comparisons and evaporation subtractions confirm the evaporative effect. The development of forward to backward comparisons and homogeneous density variations will allow us to further distinguish collisional effects from single particle dynamics in the next generation system.

\section{Scaling Law Derivation\label{sec:der}}

Here we derive the loss enhancement scaling law presented in Eqn.~\ref{eq:etaMT}, and repeated here:
\begin{equation}
\eta=\frac{3}{11} \left(\frac{d_\text{eff}E}{\sqrt{\kappa\Delta}}\right)^{8/3}.
\end{equation}
The key idea is to compare the surface areas of the loss regions with and without electric field.
There is no exact loss region where a molecule is guaranteed to spin flip, but rather its velocity and direction contribute to the Landau-Zener probability (Eqn.~\ref{eqn:lz}).
Nonetheless, for the purposes of a scaling law, we can assume the average thermal velocity $v_T$, and choose a probability threshold of $P>1/e$.
These assumptions allow us to define the loss region as the contour surface of energy $\kappa$ where
\begin{equation}
\kappa=\sqrt{2\hbar\dot{G}/\pi}=\sqrt{4\hbar v_T B'/\pi}.
\end{equation}
Here $\dot{G}$ is the rate of change in the energy gap between the trapped state and its spin flip partner, and $B'$ is the magnetic field gradient along the strong axis of the trap.

We assume that the electric field is applied parallel to the strong axis of the quadrupole trap, which makes the loss plane, as defined by $\vec{E}\perp\vec{B}$, perpendicular to this axis. This matches the geometry that has been realized in our experiment~\cite{Stuhl2013}, and is the worst case, but by no more than a constant factor of $2\sqrt{2}$ relative to other directions the electric field could have.

Before application of electric field, the $\kappa$ valued energy contour is the surface of an oblate ellipsoid of long radius $r_0=2\kappa/\mu_\text{eff}B'$.
Its area is then $2\pi\alpha\,r_0^2$, where $\alpha(e)=1+(1/e-e)\text{tanh}^{-1}(e)$ generally for eccentricity $e$, and $\alpha\sim 1.38$ for the present 2:1 ellipsoid.
When electric field is applied, the energy gap near the trap zero takes an unusual functional form.
To derive it, we first assign spatial coordinates $r$ and $z$ denoting directions within and normal to the loss plane, respectively.
Next we diagonalize the ground state hamiltonian of OH in mixed fields, see App.~A of Ref.~\cite{Stuhl2012uwave}, or similarly for another species.
Subtracting the energies of the trapped state and its spin-flip partner, and then series expanding the result yields:
\begin{equation}
\label{eqn:energy}
G = 2\mu_\text{eff}B'|z| + \beta\frac{(\mu_\text{eff}B'r/2)^3\Delta^2}{(d_\text{eff}E)^4}f(d_\text{eff}E/\Delta),
\end{equation}
plus higher order terms in $r$ and $z$.
Here $\beta=625/144=4.3$ and $f$ is a rational expression that approaches $1$ for small arguments: $f(x) = (1 + 1.28x^2)/\sqrt{1+1.44x^2}$.
The key feature, as discussed in Sec.~\ref{sec:mech}, is the cubic dependence $G$ exhibits on $r$ which leads to much more severely oblate contours.

Now we can use Eqn.~\ref{eqn:energy} to compute the surface area of the $G=\kappa$ contour.
We specialize to the regime where $d_\text{eff}E<\Delta$, so that $f(d_\text{eff}E/\Delta)\sim 1$.
The radial extent of the surface can be solved by inverting $\kappa=G|_{z=0}$:
\begin{equation}
\label{eqn:rE}
r_E = \frac{1}{\mu_\text{eff}B'}\sqrt[3]{\frac{8\kappa(d_\text{eff}E)^4}{\beta\Delta^2}}.
\end{equation}
The axial extent remains $z=\kappa/\mu_\text{eff}B'$ for all $\vec{E}$.
For large enough $E$, $r_E$ dominates over this axial extent, so that the area is effectively $2\pi r_E^2$ and the loss area enhancement becomes $\eta = r_E^2/(\alpha r_0^2)$.
Putting everything together:
\begin{equation}
\begin{split}
\label{eqn:eta}
\eta\quad =&\quad \frac{1}{\alpha}\left(\frac{1}{\mu_\text{eff}B'}\sqrt[3]{\frac{8\kappa(d_\text{eff}E)^4}{\beta\Delta^2}}\right)^2\bigg/\left(\frac{2\kappa}{\mu_\text{eff}B'}\right)^2\\
\quad=&\quad \frac{1}{\alpha\beta^{2/3}}\left(\frac{1}{2\kappa}\sqrt[3]{\frac{8\kappa(d_\text{eff}E)^4}{\Delta^2}}\right)^2\\
\quad=&\quad\frac{3}{11}\left(\frac{d_\text{eff}E}{\sqrt{\kappa\Delta}}\right)^{8/3}.
\end{split}
\end{equation}

Now we address the domain of validity of this result.
When $E$ is small, Eqn.~\ref{eqn:energy} only has a narrow range of validity, since the electric field only dominates in a very small region near the trap center.
Outside, $G$ retains a nearly linear dependence on $r$.
This means that Eqn.~\ref{eqn:rE} only holds for $E$ above some threshold.
For smaller $E$, $r_E$ will simply not be significantly perturbed from its zero electric field value of $r_0=2\kappa/\mu_\text{eff}B'$.
The implication for the enhancement factor in Eqn.~\ref{eqn:eta} is simply that it is only valid when it predicts an enhancement significantly greater than unity.
In other words, Eqn.~\ref{eqn:eta} holds when $d_\text{eff}E>1.6\!\cdot\!\sqrt{\kappa\Delta}$, but below this $\eta$ gradually returns to unity.
Eventually when $d_\text{eff}E>\Delta$, the factor of $f(d_\text{eff}E/\Delta)$ in Eqn.~\ref{eqn:energy} is better approximated by $1.1\!\cdot\!d_\text{eff}E/\Delta$, which leads to the modification $\eta=0.26\!\cdot\!(d_\text{eff}E)^2/\kappa^{4/3}\Delta^{2/3}$.
Thus for these larger E-fields, the enhancement factor reduces in its dependence on electric field from order $8/3$ to order $2$.
At this point, the loss is typically too large for trapping, see Tab.~\ref{tab:rates}.

\bibliographystyle{apsrev4-1_no_Arxiv}
\bibliography{LF16145_Reens_MSFL}

\end{document}